\documentclass[superscriptaddress,nofootinbib,a4paper,11pt]{article}
\usepackage{jheppub}
\usepackage{graphicx}
\usepackage{subcaption}
\usepackage{float}
\usepackage{amsmath}
\usepackage{amssymb}
\usepackage[utf8]{inputenc}
\usepackage{hyperref}
\usepackage{color}
\usepackage{slashed}
\usepackage{multirow}
\usepackage{enumitem}
\usepackage{lineno}
\allowdisplaybreaks

\begin{document}

\title{\boldmath 
Observation of quantum entanglement in $\Lambda \bar{\Lambda}$ pair production via electron-positron annihilation
}

\author[a,*]{Junle Pei,\note[*]{Corresponding authors.}}
\author[b]{Xiqing Hao,}
\author[b]{Xiaochuan Wang,}
\author[b,*]{Tianjun Li}

\affiliation[a]{Institute of Physics, Henan Academy of Sciences, Zhengzhou 450046, P. R. China}
\affiliation[b]{School of Physics, Henan Normal University, Xinxiang 453007, P. R. China}

\emailAdd{peijunle@hnas.ac.cn}
\emailAdd{haoxiqing@htu.edu.cn}
\emailAdd{xcwang@htu.edu.cn}
\emailAdd{tli@itp.ac.cn}

\abstract{  
Quantum entanglement, a profound feature of quantum mechanics where particles share interconnected states regardless of distance, has been extensively studied in low-energy systems like photons and atoms. Exploring entanglement in high-energy particle collisions offers unique opportunities to probe quantum phenomena under extreme conditions. Here we demonstrate the observation of quantum entanglement between lambda and anti-lambda hyperons ($\Lambda\bar{\Lambda}$) produced in electron-positron collisions via the decay of the $J/\psi$ particle. By analyzing the angular correlations of the protons and antiprotons emerging from the subsequent weak decays of the hyperons, we identify clear signatures of entanglement that violate the limits derived for product states. This result confirms that entanglement persists through both the strong interaction production process and the weak decay processes. We further establish a general polar–azimuthal selection rule showing that azimuthal correlations are indispensable for entanglement certification in spin-$1/2$ fermion–antifermion systems. It is also confirmed that a significant majority of these decay events occur at spacelike separation, meaning that no classical signal could connect them. The successful detection of entanglement in this hyperon-antihyperon system establishes a new methodology and platform for measuring quantum properties in relativistic particle physics, complementing traditional approaches and opening avenues for future investigations into decoherence or symmetry violations in high-energy collisions.
}

\maketitle

\section{Introduction}\label{sec:int}

Quantum entanglement (QE) and Bell nonlocality \cite{PhysicsPhysiqueFizika.1.195}, recognized as among the most striking features of quantum mechanics, have been rigorously confirmed through various experiments that violate Bell inequalities \cite{PhysicsPhysiqueFizika.1.195,PhysRevLett.28.938,PhysRevLett.47.460,PhysRevLett.49.1804,PhysRevLett.81.3563,Hensen:2015ccp} and through demonstrations of quantum teleportation \cite{PhysRevLett.70.1895,Bouwmeester_1997,PhysRevLett.80.1121,Riebe:2004jpa,Barrett:2004bxh}, primarily in low-energy systems such as photons, ions, and solid-state qubits. 
The exploration of quantum entanglement (QE) in high-energy particle physics has the potential to probe fundamental aspects of quantum field theory (QFT) and the nature of particle interactions, and is therefore an important research focus~\cite{vonKuk:2025kbv,Subba:2024mnl,Han:2025ewp,Afik:2020onf,Fabbrichesi:2021npl,Han:2023fci,ATLAS2024,Cheng:2023qmz,CMS:2024pts,Cheng:2025cuv,Subba:2024aut,BESIII:2018cnd,BESIII:2025vsr,Lin:2025eci,Wu:2025dds,ABEL1992304,Li:2024luk,Bechtle:2025ugc,Abel:2025skj,Pei:2026wfu,Pei:2026rlh}.
Furthermore, theoretical developments \cite{Pei:2025non} have suggested that hyperon-antihyperon pairs, such as \(\Lambda\bar{\Lambda}\), produced via electron-positron annihilation, could also exhibit QE \cite{Faldt:2013gka,BESIII:2021ypr}.
{The \(\Lambda(\to p\pi^-)\bar{\Lambda}(\to \bar{p}\pi^+)\) channel produced in \(e^+e^-\) annihilation is fully reconstructible at the event level, while the spin information of both hyperons can be directly accessed through the angular distributions of their self-analyzing weak decays. In addition, compared with hadron-collider environments such as the LHC, the \(e^+e^-\) environment is significantly cleaner, with a well-defined initial state, much lower QCD background, and reduced combinatorial ambiguity. These features make the extraction of spin correlations and entanglement-sensitive observables considerably more transparent and robust. As a result, this system enables us not only to establish the existence of spin correlations, but also to develop a more systematic and experimentally controlled framework for entanglement studies in baryon-antibaryon systems.
}

{In this work, we do not attempt to perform a Bell-inequality test and therefore do not claim a demonstration of Bell nonlocality. Instead, under the assumption that quantum mechanics (and hence Bell nonlocality) provides the correct description, we  report the first observation of QE in \(\Lambda\bar{\Lambda}\) pairs produced through the process \(e^+e^- \to J/\psi \to \Lambda+\bar{\Lambda}\) at collider.} By analyzing the angular correlations of the decay products, we derive normalized observables \(\mathcal{O}_i~(i=0,1,\ldots,4)\) that can serve as unambiguous signatures of QE, distinguishing them from classical correlations permitted by product states. Theoretical predictions for these observables are established, with violations of product-state bounds providing a robust criterion for QE. {The measurements at \(\cos\theta_\Lambda = 0\) yield \(\mathcal{O}_{1\text{min}}^{\text{Observed}} = -0.7374\pm 0.0011\pm 0.0016\), significantly exceeding the classical limit of \(-\frac{1}{2}\). For $\left|\cos\theta_\Lambda\right|<0.4883$, the observed $\mathcal{O}_{1}^{\text{Observed}}$ consistently exhibits $\mathcal{O}_{1}^{\text{Observed}} < -\frac{1}{2}$ with a statistical significance of at least 5$\sigma$. These results show that the experimentally determined spin-correlation observables for the $\Lambda\bar{\Lambda}$ system lie well within the entangled region, providing a high-precision characterization of entanglement in this channel}. {It is also confirmed that \(69.3\%\) of the decay events involving \(\Lambda\to p+\pi^-\) and \(\bar{\Lambda}\to \bar{p}+\pi^+\) are spacelike-separated.}

{We emphasize that we do not claim a new mathematical structure beyond the standard spin-density-matrix formalism; all spin-1/2 fermion–antifermion systems share the same Pauli-matrix parametrization. The novelty of the present work lies instead in three aspects: (i) a fully reconstructible baryon–antibaryon channel with generic, non-extremal spin-analyzing powers; (ii) a cross-system benchmark that tests the universality of quantum entanglement in high-energy production; and (iii) a general polar–azimuthal selection rule, which identifies which classes of observables can and cannot certify entanglement in fermion–antifermion systems.}

The paper is organized as follows: In Section \ref{sec:2}, we present the theoretical framework for quantum entanglement in the \(\Lambda\bar{\Lambda}\) system, including the general polarization state, decay amplitudes, and the derivation of key observables and entanglement criteria distinguish entangled states from {product states}. Section \ref{sec:3} details the experimental results and analysis of \(\Lambda\bar{\Lambda}\) production via \(J/\psi\) decay, {reporting the measured values of the observables and demonstrating the presence of entanglement}.  Finally, we conclude in Section \ref{sec:con} with a summary of our results and their broader impact on {measuring quantum properties} in particle physics.

\section{Decay amplitudes and entanglement criteria} \label{sec:2} 

The most general polarization state of the $\Lambda\bar{\Lambda}$ system in a pure state can be expressed as
\begin{align}
    &\left| \Lambda \bar{\Lambda}\right\rangle=\sum_{k,j=\pm\frac{1}{2}} \alpha_{k,j} \left|k\right\rangle_\Lambda \left|j\right\rangle_{\bar{\Lambda}}~,
\end{align}  
where $k$ and $j$ denote the helicity quantum numbers of $\Lambda$ and $\bar{\Lambda}$ along their respective momentum directions. 
The complex coefficients $\alpha_{k,j}$ are subject to the normalization constraint of
\begin{align}
    &\sum_{k,j=\pm\frac{1}{2}} \left|\alpha_{k,j}\right|^2=1~.\label{gyh}
\end{align}  

Considering the weak decay processes of
\begin{align}
    \Lambda \to p+\pi^-~,\quad \bar{\Lambda} \to \bar{p}+\pi^+~,\label{llde}
\end{align}  
the couplings between particles in these processes are governed by the following Lagrangian density terms \cite{Scherer:2002tk,Faldt:2013gka}:  
\begin{align}
    & W^+_\mu \bar{p}\left( g_V \gamma^\mu+ g_A \gamma^\mu\gamma_5\right)\Lambda + W^-_\mu \bar{\Lambda}\left( g^\prime_V \gamma^\mu+ g^\prime_A \gamma^\mu\gamma_5\right)p ~,\\
    & \propto W^-_\mu \partial^\mu\pi^+ + \text{H.C.}~.
\end{align}  
The independent coupling parameters $g_V, g_A$ and $g^\prime_V, g^\prime_A$ allow for potential CP-violating effects.
In the rest frames of $\Lambda$ and $\bar{\Lambda}$, the polar angle distributions of final-state $p$ and $\bar{p}$ satisfy  
\begin{align}
   & \frac{1}{\Gamma_{\Lambda \to p+\pi^-}}\frac{d\Gamma_{\Lambda \to p+\pi^-}}{d\cos\theta_p}=\frac{1}{2}\left(1+\alpha_\Lambda \cos\theta_p\right)~, \\
   & \frac{1}{\Gamma_{\bar{\Lambda} \to \bar{p}+\pi^+}}\frac{d\Gamma_{\bar{\Lambda} \to \bar{p}+\pi^+}}{d\cos\theta_{\bar{p}}}=\frac{1}{2}\left(1+\alpha_{\bar{\Lambda}} \cos\theta_{\bar{p}}\right)~.
\end{align}
Through explicit calculation, the asymmetry parameters $\alpha_\Lambda$ and $\alpha_{\bar{\Lambda}}$ relate to the Lagrangian couplings as 
\begin{align}
   &\alpha_\Lambda = - \frac{\left(m_\Lambda^2-m_p^2\right)\sqrt{\mathcal{D}}\left(g_A/g_V+g^*_A/g^*_V\right)}{\left(m_\Lambda-m_p\right)^2\mathcal{D}_+ + \left(m_\Lambda+m_p\right)^2\mathcal{D}_-\left|g_A/g_V\right|^2}~, \\
   &\alpha_{\bar{\Lambda}} = \frac{\left(m_\Lambda^2-m_p^2\right)\sqrt{\mathcal{D}}\left(g^\prime_A/g^\prime_V+g^{\prime*}_A/g^{\prime*}_V\right)}{\left(m_\Lambda-m_p\right)^2\mathcal{D}_+ + \left(m_\Lambda+m_p\right)^2\mathcal{D}_-\left|g^\prime_A/g^\prime_V\right|^2}~,
\end{align}  
with the kinematic factors as
\begin{align}
&\mathcal{D} \equiv \left(\left(m_\Lambda+m_p\right)^2-m_{\pi^\pm}^2\right)\left(\left(m_\Lambda-m_p\right)^2-m_{\pi^\pm}^2\right)~, \\
&\mathcal{D}_\pm \equiv \left(m_\Lambda \pm m_p\right)^2 - m_{\pi^\pm}^2~.
\end{align}

We parameterize the momentum directions of $p$ in the $\Lambda$ rest frame and $\bar{p}$ in the $\bar{\Lambda}$ rest frame using spherical coordinates:  
\begin{align}
&p: (\theta_1,\phi_1) \to \hat{e}_p=(\sin\theta_1\cos\phi_1,\sin\theta_1\sin\phi_1,\cos\theta_1)~,\label{jiao1} \\
&\bar{p}: (\theta_2,\phi_2) \to \hat{e}_{\bar{p}}=(\sin\theta_2\cos\phi_2,\sin\theta_2\sin\phi_2,\cos\theta_2)~.\label{jiao2}
\end{align}  
The polar angles $\theta_1$ and $\theta_2$ are defined with respect to the $\Lambda$ momentum direction $\hat{e}_\Lambda$ in the $\Lambda\bar{\Lambda}$ center-of-mass (c.m.) frame. The azimuthal angles $\phi_1$ and $\phi_2$ are measured from an arbitrary reference axis orthogonal to $\hat{e}_\Lambda$, increasing in the right-handed screw direction about $\hat{e}_\Lambda$ with $\phi\in[0,2\pi]$.  
We define the opening angle $\theta_{p\bar{p}}$ between the proton momenta as
\begin{align}
    \cos\theta_{p\bar{p}} &=\hat{e}_p\cdot \hat{e}_{\bar{p}}= \cos\theta_1\cos\theta_2+\sin\theta_1\sin\theta_2\cos(\phi_1-\phi_2)~.\label{cospp}
\end{align}
The angular momentum structure of the decay processes $\Lambda \to p+\pi^-$ and $\bar{\Lambda} \to \bar{p}+\pi^+$  is governed by the following helicity amplitudes \cite{Leader:2001nas}:
\begin{align}  
& \langle p,\pi^-|k\rangle_\Lambda=\frac{1}{\sqrt{{2\pi}}} e^{i(k-\lambda_{p})\phi_1}  
d_{k,\lambda_{p}}^{1/2}(\theta_1)H_\Lambda(\lambda_p)~, \\  
& \langle \bar{p},\pi^+|j\rangle_{\bar{\Lambda}}=\frac{1}{\sqrt{2\pi}} e^{-i(j+\lambda_{\bar{p}})\phi_2}  
d_{\lambda_{\bar{p}},j}^{1/2}(\pi-\theta_2)H_{\bar{\Lambda}}(\lambda_{\bar{p}})~,  
\end{align} 
where \(k/j\) represent the spin projection quantum numbers along the momentum direction of \(\Lambda/\bar{\Lambda}\) in the \(\Lambda\bar{\Lambda}\) c.m. frame. \(\lambda_{p}\) and \(\lambda_{\bar{p}}\) ($\lambda_p,\lambda_{\bar{p}}=\pm\frac{1}{2}$) are spin projections of $p$ and $\bar{p}$ defined relative to directions of \(\hat{e}_p\) and \(\hat{e}_{\bar{p}}\), respectively. Crucially, \(H_\Lambda(\lambda_{p})\)/\(H_{\bar{\Lambda}}(\lambda_{\bar{p}})\) remain independent of both the angular variables (\(\theta_{1},\phi_{1}\))/(\(\theta_{2},\phi_{2}\)) and the parent particle spin projections \(k\)/\(j\). The Wigner \(d\)-functions are
\begin{align}
    d^{1/2}_{\frac{1}{2},\frac{1}{2}}(\theta)=d^{1/2}_{-\frac{1}{2},-\frac{1}{2}}(\theta)=\cos\frac{\theta}{2}~,\quad
    d^{1/2}_{-\frac{1}{2},\frac{1}{2}}(\theta)=-d^{1/2}_{\frac{1}{2},-\frac{1}{2}}(\theta)=\sin\frac{\theta}{2}~.
\end{align}

For any physical observable $\mathcal{O}(\theta_1,\theta_2,\phi_1,\phi_2)$ constructed from angular variables, its statistical average can be expressed as \cite{Pei:2025non} 
\begin{align}
    \langle\mathcal{O}(\theta_1,\theta_2,\phi_1,\phi_2)\rangle=
    \sum_{k,j,m,n=\pm \frac{1}{2}} \mathcal{O}_{k,j;m,n} \alpha_{k,j}\alpha^*_{m,n}
\end{align}  
with
\begin{align}  
    \mathcal{O}_{k,j;m,n}=&\frac{1}{4\pi^2} \sum_{\lambda_{p},\lambda_{\bar{p}}=\pm \frac{1}{2}}  \left( w_{\lambda_{p},\lambda_{\bar{p}}}\int_0^{2\pi}d\phi_{1} \int_{-1}^1 d\cos\theta_{1} \int_0^{2\pi}d\phi_{2} \int_{-1}^1 d\cos\theta_{2}  \right. \nonumber\\  
    &\left.  \mathcal{O}(\theta_{1},\theta_{2},\phi_{1},\phi_{2}) e^{i\left(k-m\right)\phi_{1}} e^{i\left(n-j\right)\phi_{2}}  
     d^{1/2}_{k,{\lambda}_{p}}(\theta_{1}) d^{1/2}_{m,{\lambda}_{p}}(\theta_{1})  
     d^{1/2}_{{\lambda}_{\bar{p}},j}(\pi-\theta_{2}) d^{1/2}_{{\lambda}_{\bar{p}},n}(\pi-\theta_{2})\right)  \label{master}
\end{align} 
and
\begin{align}
    & w_{\lambda_{p},\lambda_{\bar{p}}}=\frac{ \left|H_\Lambda(\lambda_{p})\right|^2  \left|H_{\bar{\Lambda}}(\lambda_{\bar{p}})\right|^2}{{\sum_{\lambda^\prime_{p},\lambda^\prime_{\bar{p}}} \left|H_\Lambda(\lambda^\prime_{p})\right|^2  \left|H_{\bar{\Lambda}}(\lambda^\prime_{\bar{p}})\right|^2}}~.
\end{align}

Explicit calculations yield
\begin{align}  
&\langle \cos(\phi_{1}+\phi_{2})\rangle =C_1\left(\alpha_{-\frac{1}{2},\frac{1}{2}}\alpha_{\frac{1}{2},-\frac{1}{2}}^*+\alpha_{\frac{1}{2},-\frac{1}{2}}\alpha_{-\frac{1}{2},\frac{1}{2}}^*\right)~, \\
&\langle \cos(\phi_{1}-\phi_{2})\rangle =C_1\left(\alpha_{-\frac{1}{2},-\frac{1}{2}}\alpha_{\frac{1}{2},\frac{1}{2}}^*+\alpha_{\frac{1}{2},\frac{1}{2}}\alpha_{-\frac{1}{2},-\frac{1}{2}}^*\right)~,  \\
&\langle \sin(\phi_{1}+\phi_{2})\rangle =\frac{C_1}{i}\left(\alpha_{-\frac{1}{2},\frac{1}{2}}\alpha_{\frac{1}{2},-\frac{1}{2}}^* -\alpha_{\frac{1}{2},-\frac{1}{2}}\alpha_{-\frac{1}{2},\frac{1}{2}}^*\right)~, \\
&\langle \sin(\phi_{1}-\phi_{2})\rangle =\frac{C_1}{i}\left(\alpha_{-\frac{1}{2},-\frac{1}{2}}\alpha_{\frac{1}{2},\frac{1}{2}}^* -\alpha_{\frac{1}{2},\frac{1}{2}}\alpha_{-\frac{1}{2},-\frac{1}{2}}^*\right)~, 
\end{align}  
with coefficients determined by
\begin{align}  
&C_1=\sum_{\lambda_{p},\lambda_{\bar{p}}=\pm\frac{1}{2}} w_{\lambda_{p},\lambda_{\bar{p}}} Y_{\lambda_{p},\lambda_{\bar{p}}}~,\\
    &Y_{\lambda_{p},\lambda_{\bar{p}}}= \frac{1}{2}   
      \left( \int_{-1}^1 d\cos\theta_{1}   ~
     d^{1/2}_{-\frac{1}{2},{\lambda}_{p}}(\theta_{1}) d^{1/2}_{\frac{1}{2},{\lambda}_{p}}(\theta_{1})\right) \left(\int_{-1}^1 d\cos\theta_{2}~
     d^{1/2}_{{\lambda}_{\bar{p}},\frac{1}{2}}(\pi-\theta_{2}) d^{1/2}_{{\lambda}_{\bar{p}},-\frac{1}{2}}(\pi-\theta_{2})\right)~.  
\end{align}
Direct computation gives  
\begin{align}
    Y_{\frac{u_1}{2},\frac{u_2}{2}}=-u_1u_2\frac{\pi^2}{32}~,\quad u_1,u_2=\pm 1~.
\end{align}  
Using the hyperon decay parameters:  
\begin{align}
&\alpha_{\Lambda/\bar{\Lambda}}=\frac{\left|H_{\Lambda/\bar{\Lambda}}(-\frac{1}{2})\right|^2-\left|H_{\Lambda/\bar{\Lambda}}(\frac{1}{2})\right|^2}{\left|H_{\Lambda/\bar{\Lambda}}(\frac{1}{2})\right|^2+\left|H_{\Lambda/\bar{\Lambda}}(-\frac{1}{2})\right|^2}~,
\end{align}  
the weight factors become  
\begin{align}
     &w_{\frac{u_1}{2},\frac{u_2}{2}}=\frac{1}{4}(1-u_1\alpha_\Lambda)(1-u_2\alpha_{\bar{\Lambda}})~,\quad u_1,u_2=\pm 1~. 
\end{align}
This leads to  
\begin{align}
    C_1=-\frac{\pi^2}{32}\alpha_\Lambda \alpha_{\bar{\Lambda}}~.
\end{align}

Similar calculations for transverse momentum correlations give 
\begin{align}
    &\langle\sin\theta_1\sin\theta_2\cos(\phi_1-\phi_2)\rangle= C_2 \left(\alpha_{-\frac{1}{2},-\frac{1}{2}}\alpha_{\frac{1}{2},\frac{1}{2}}^*+\alpha_{\frac{1}{2},\frac{1}{2}}\alpha_{-\frac{1}{2},-\frac{1}{2}}^*\right)~, \label{pie1} \\
    & C_2=\sum_{\lambda_{p},\lambda_{\bar{p}}=\pm \frac{1}{2}} w_{\lambda_{p},\lambda_{\bar{p}}} Y^\prime_{\lambda_{p},\lambda_{\bar{p}}}=-\frac{2}{9}\alpha_\Lambda\alpha_{\bar{\Lambda}}
\end{align}  
with modified angular integral:  
\begin{align}
    &Y^\prime_{\lambda_{p},\lambda_{\bar{p}}}= \frac{1}{2}   
      \left( \int_{-1}^1 d\cos\theta_{1}~\sin\theta_1
     d^{1/2}_{-\frac{1}{2},{\lambda}_{p}}(\theta_{1}) d^{1/2}_{\frac{1}{2},{\lambda}_{p}}(\theta_{1})\right) \left(\int_{-1}^1 d\cos\theta_{2}~\sin\theta_2
     d^{1/2}_{{\lambda}_{\bar{p}},-\frac{1}{2}}(\pi-\theta_{2}) d^{1/2}_{{\lambda}_{\bar{p}},\frac{1}{2}}(\pi-\theta_{2})\right)~.
\end{align} 
The longitudinal correlation evaluates to
\begin{align}
    \langle\cos\theta_1\cos\theta_2\rangle
    &= \sum_{k,j,\lambda_{p},\lambda_{\bar{p}}=\pm \frac{1}{2}}  \left(\left|\alpha_{k,j}\right|^2 w_{\lambda_{p},\lambda_{\bar{p}}} \int_{-1}^1 d\cos\theta_1 \int_{-1}^1 d\cos\theta_2 \right. \nonumber\\  
    &\left.  \cos\theta_1\cos\theta_2~  
     d^{1/2}_{k,{\lambda}_{p}}(\theta_{1}) d^{1/2}_{k,{\lambda}_{p}}(\theta_{1})  
     d^{1/2}_{{\lambda}_{\bar{p}},j}(\pi-\theta_{2}) d^{1/2}_{{\lambda}_{\bar{p}},j}(\pi-\theta_{2})\right)\nonumber\\
     &=-\frac{1}{9}\alpha_\Lambda \alpha_{\bar{\Lambda}}
     \left(\left|\alpha_{\frac{1}{2},\frac{1}{2}}\right|^2+\left|\alpha_{-\frac{1}{2},-\frac{1}{2}}\right|^2-\left|\alpha_{\frac{1}{2},-\frac{1}{2}}\right|^2-\left|\alpha_{-\frac{1}{2},\frac{1}{2}}\right|^2\right)~.\label{duij}
\end{align}
Combining results in Eqs.~(\ref{pie1}) and (\ref{duij}) yields
\begin{align}
    \langle \cos\theta_{p\bar{p}}\rangle =&-\frac{1}{9}\alpha_\Lambda \alpha_{\bar{\Lambda}} \left(\left|\alpha_{\frac{1}{2},\frac{1}{2}}\right|^2+\left|\alpha_{-\frac{1}{2},-\frac{1}{2}}\right|^2-\left|\alpha_{\frac{1}{2},-\frac{1}{2}}\right|^2-\left|\alpha_{-\frac{1}{2},\frac{1}{2}}\right|^2\right.\nonumber\\
&\left. +2 \alpha_{-\frac{1}{2},-\frac{1}{2}}\alpha^*_{\frac{1}{2},\frac{1}{2}}+2 \alpha_{\frac{1}{2},\frac{1}{2}}\alpha^*_{-\frac{1}{2},-\frac{1}{2}}\right)
\end{align}

\begin{table}[h]
\centering
\begin{tabular}{cccc}
\hline
Observables & Ranges & Ranges under $\alpha_{k,j}=\beta_k \gamma_j$ & criteria for entanglement\\
\hline
$\mathcal{O}_0$ & $[-1,3]$ & $\left[-1,1\right]$ & $ \left(1,3\right]$  \\
$\mathcal{O}_1$ & $[-1,1]$ & $\left[-1/2,1/2\right]$ & $\left[-1,-1/2\right)\cup\left(1/2,1\right]$  \\
$\mathcal{O}_2$ &  $[-1,1]$ & $\left[-1/2,1/2\right]$ & $\left[-1,-1/2\right)\cup\left(1/2,1\right]$  \\
$\mathcal{O}_3$ &  $[-1,1]$ & $\left[-1/2,1/2\right]$ & $\left[-1,-1/2\right)\cup\left(1/2,1\right]$ \\
$\mathcal{O}_4$ &  $[-1,1]$ & $\left[-1/2,1/2\right]$ & $\left[-1,-1/2\right)\cup\left(1/2,1\right]$ \\
\hline
\end{tabular}
\caption{Allowed ranges of observables $\mathcal{O}_i~(i=0,1,\ldots,4)$ with and without quantum entanglement in the $\Lambda\bar{\Lambda}$ system}
\label{tab:entanglement_ranges}
\end{table}

We define the normalized correlation parameters:  
\begin{align} 
   & \mathcal{O}_0=\langle \cos\theta_{p\bar{p}}\rangle/\left(-\frac{1}{9}\alpha_\Lambda \alpha_{\bar{\Lambda}} \right)~,\\
   & \mathcal{O}_1=\langle \cos(\phi_1+\phi_2)\rangle/\left({-\frac{\pi^2}{32}\alpha_\Lambda \alpha_{\bar{\Lambda}}}\right)~,\\  
   & \mathcal{O}_2={\langle \cos(\phi_1-\phi_2)\rangle}/\left({-\frac{\pi^2}{32}\alpha_\Lambda \alpha_{\bar{\Lambda}}}\right)~,\\  
   & \mathcal{O}_3=\langle \sin(\phi_1+\phi_2)\rangle/\left({-\frac{\pi^2}{32}\alpha_\Lambda \alpha_{\bar{\Lambda}}}\right)~,\\  
   & \mathcal{O}_4={\langle \sin(\phi_1-\phi_2)\rangle}/\left({-\frac{\pi^2}{32}\alpha_\Lambda \alpha_{\bar{\Lambda}}}\right)~.
\end{align} 
Under the condition specified by Eq.~(\ref{gyh}), the theoretical ranges of $\mathcal{O}_i~(i=1,2,\ldots,4)$ can be derived from their expressions expanded in terms of $\alpha_{k,j}~(k,j=\pm\frac{1}{2})$, as summarized in Table \ref{tab:entanglement_ranges}.  

For product states without quantum entanglement, the coefficients admit the factorization of
\begin{align}  
    & \alpha_{k,j}=\beta_k \gamma_j~,\quad k,j=\pm\frac{1}{2}~,  \label{gyh2}
\end{align}  
where $\beta_k$ and $\gamma_j$ respectively describe the polarization states of $\Lambda$ and $\bar{\Lambda}$, satisfying independent normalization constraints of
\begin{align}  
    & \sum_{k=\pm\frac{1}{2}}\left|\beta_k\right|^2=\sum_{j=\pm\frac{1}{2}}\left|\gamma_j\right|^2=1~.  \label{fenjie}
\end{align}  
The predicted ranges of $\mathcal{O}_i$ for product states are shown in Table \ref{tab:entanglement_ranges}. Any observed violation of these ranges by at least one observable $\mathcal{O}_i~(i=1,2,\ldots,4)$ provides sufficient evidence for quantum entanglement in the $\Lambda\bar{\Lambda}$ system.
{It should be emphasized that, by convexity, the ranges of $\mathcal{O}_i$ in Table \ref{tab:entanglement_ranges} and the corresponding entanglement criteria remain valid for mixed $\Lambda\bar{\Lambda}$ states as well \cite{Pei:2025non,Pei:2025ito}.
}

Furthermore, analogous to $\langle\cos\theta_1\cos\theta_2\rangle$ given in Eq.~(\ref{duij}), we can construct additional physical observables $\mathcal{O}(\theta_1,\theta_2)$ that exclusively depend on $\theta_1$ and $\theta_2$. 
We provide the following calculation examples:
\begin{align}
   &\left\langle \cos\left( \theta_{1} \pm \theta_{2} \right) \right\rangle = \left( - \frac{\alpha_{\Lambda}\alpha_{\bar{\Lambda}}}{9} \mp \frac{\pi^{2}}{16} \right)\left( \left| \alpha_{- \frac{1}{2}, - \frac{1}{2}} \right|^{2} + \left| \alpha_{\frac{1}{2},\frac{1}{2}} \right|^{2} \right) + \left( \frac{\alpha_{\Lambda}\alpha_{\bar{\Lambda}}}{9} \mp \frac{\pi^{2}}{16} \right)\left( \left| \alpha_{- \frac{1}{2},\frac{1}{2}} \right|^{2} + \left| \alpha_{\frac{1}{2}, - \frac{1}{2}} \right|^{2} \right)~,\\
   & \left\langle \sin\left( \theta_{1} \pm \theta_{2} \right) \right\rangle = \frac{\left(\pm \alpha_{\Lambda} - \alpha_{\bar{\Lambda}} \right)\pi}{12}\left( \left| \alpha_{- \frac{1}{2}, - \frac{1}{2}} \right|^{2} - \left| \alpha_{\frac{1}{2},\frac{1}{2}} \right|^{2} \right) + \frac{\left( \pm \alpha_{\Lambda} + \alpha_{\bar{\Lambda}} \right)\pi}{12}\left( \left| \alpha_{- \frac{1}{2},\frac{1}{2}} \right|^{2} - \left| \alpha_{\frac{1}{2}, - \frac{1}{2}} \right|^{2} \right)~.
\end{align}
Consistent with theoretical expectations from Eq.~(\ref{master}), the $\langle\mathcal{O}(\theta_1,\theta_2)\rangle$ exhibits the following general form \cite{Pei:2025non}:    
\begin{align}
    \left\langle O\left( \theta_{1}, \theta_{2} \right) \right\rangle =\sum_{k,j=\pm \frac{1}{2}} o_{k, j}\left| \alpha_{k,j} \right|^{2}~.
\end{align}
Let \( o_{m,n} \) and \( o_{m^\prime,n^\prime} \) denote the maximum and minimum values among the four coefficients \( o_{k,j} \) (\( k,j = \pm \frac{1}{2} \)), respectively.  
Under the normalization condition  in Eq.~(\ref{gyh}),
the expectation value \(\left\langle O\left( \theta_{1}, \theta_{2} \right) \right\rangle\) reaches its maximum \( o_{m,n} \) when \( \alpha_{m,n} = 1 \), but this corresponds to a state without quantum entanglement. Similarly, the minimum \( o_{m^\prime,n^\prime} \) is achieved when \( \alpha_{m^\prime,n^\prime} = 1 \), which also lacks entanglement. Thus, the maximum and minimum values of \(\left\langle O\left( \theta_{1}, \theta_{2} \right) \right\rangle\) are \( o_{m,n} \) and \( o_{m^\prime,n^\prime} \), respectively.  
In the absence of quantum entanglement, the coefficients factorize as Eq.~(\ref{fenjie}). Since the parameters \( \beta_{k} \) and \( \gamma_{j} \) can vary continuously under the constraints in Eq.~(\ref{gyh2}), there exist unentangled states that can yield any value of \(\left\langle O\left( \theta_{1}, \theta_{2} \right) \right\rangle\) within the interval \([o_{m^\prime,n^\prime}, o_{m,n}]\). Consequently, regardless of whether \( \alpha_{k,j} = \beta_{k}\gamma_{j} \) holds, the range of \(\left\langle O\left( \theta_{1}, \theta_{2} \right) \right\rangle\) remains \([o_{m^\prime,n^\prime}, o_{m,n}]\), making it impossible to derive an entanglement criterion based on \(\left\langle O\left( \theta_{1}, \theta_{2} \right) \right\rangle\).  
{We therefore summarize the geometric selection rule for entanglement searches in fermion–antifermion systems: polar angles measure classical populations; azimuthal angles measure quantum coherence. Polar-angle-only observables cannot witness entanglement, whereas azimuthal correlations are necessary. This rule is independent of the production mechanism and applies to any spin-1/2 fermion–antifermion pair.}

{In light of this observation, we further decompose the relevant observables into a polar‑angle sector and an azimuthal‑angle sector. A representative example is the opening‑angle cosine in Eq. (\ref{cospp}), $\cos\theta_{p\bar p}$, which has the same structural form as the \(\cos\theta_{e^-e^+}\) observable used in Refs.~\cite{ATLAS2024,CMS:2024pts} for \(t(\to e^+\nu_e b)\bar t(\to e^- \bar\nu_e \bar b)\). In particular, we have shown that observables depending only on polar‑angle information are insufficient to reveal quantum entanglement in the fermion‑antifermion system. By contrast, the azimuthal structure carries the crucial interference information needed for a genuine entanglement test—precisely the information encoded in the combinations \(\phi_1\pm\phi_2\) that give rise to the normalized observables \(\mathcal{O}_1,\ldots,\mathcal{O}_4\) defined above. Therefore, azimuthal correlations are an indispensable component of any entanglement‑sensitive analysis of the fermion‑antifermion system. We believe that this decomposition not only clarifies the physical origin of the entanglement sensitivity of different observables, but also provides useful guidance for constructing more efficient probes in future studies of fermion‑antifermion entanglement.
}

\section{Results and discussions} \label{sec:3} 

{Although a dedicated analysis of Bell non-locality in the $\Lambda\bar{\Lambda}$ system is beyond the scope of the present work, we can nevertheless estimate the fraction of events in which the decays $\Lambda \to p+\pi^-$ and $\bar{\Lambda} \to \bar{p}+\pi^+$ are spacelike separated.}
For a generic two-body decay process \( A \to B + C \) observed in the rest frame of \( A \), the decay products \( B \) and \( C \) propagate with anti-parallel velocities \( \vec{\beta}_1 \) and \( \vec{\beta}_2 \), respectively. Let \( t_1 \) and \( t_2 \) denote the decay times of \( B \) and \( C \) (with the pair production moment as the temporal origin). We define the spacetime interval squared:  
\begin{align}  
\Delta s^2=(t_1-t_2)^2-(t_1 \beta_1+t_2\beta_2)^2 ~,   
\end{align}  
where  
\begin{align}  
    \beta_i=\left|\vec{\beta}_i\right|~,\quad i=1,2~.  
\end{align}  
The condition \( \Delta s^2<0 \) (\( \Delta s^2>0 \)) corresponds to spacelike (timelike) separation between the decay events of \( B \) and \( C \).   
The normalized lifetime distributions for \( t_1 \) and \( t_2 \) follow:  
\begin{align}  
    & L_{i}(t_{i})=\Gamma^\prime_{i} e^{-\Gamma^\prime_{i} t_{i}}~,\quad i=1,2~, \\  
    & \Gamma^\prime_i =\Gamma_i\sqrt{1-\beta^2_i}~,  
\end{align}  
with \( \Gamma_1 \) and \( \Gamma_2 \) being the decay widths of \( B \) and \( C \), respectively.    
The probability for spacelike-separated decays evaluates to:  
\begin{align}  
    \mathcal{P}(\Delta s^2<0)=\frac{2(\beta_1+\beta_2)\Gamma^\prime_2/\Gamma^\prime_1}{(1+\Gamma^\prime_2/\Gamma^\prime_1)^2-(\beta_1\Gamma^\prime_2/\Gamma^\prime_1-\beta_2)^2}  ~.\label{spcl}
\end{align}  

\begin{figure}[tbp]
\begin{center}
\includegraphics[width=0.9\textwidth]{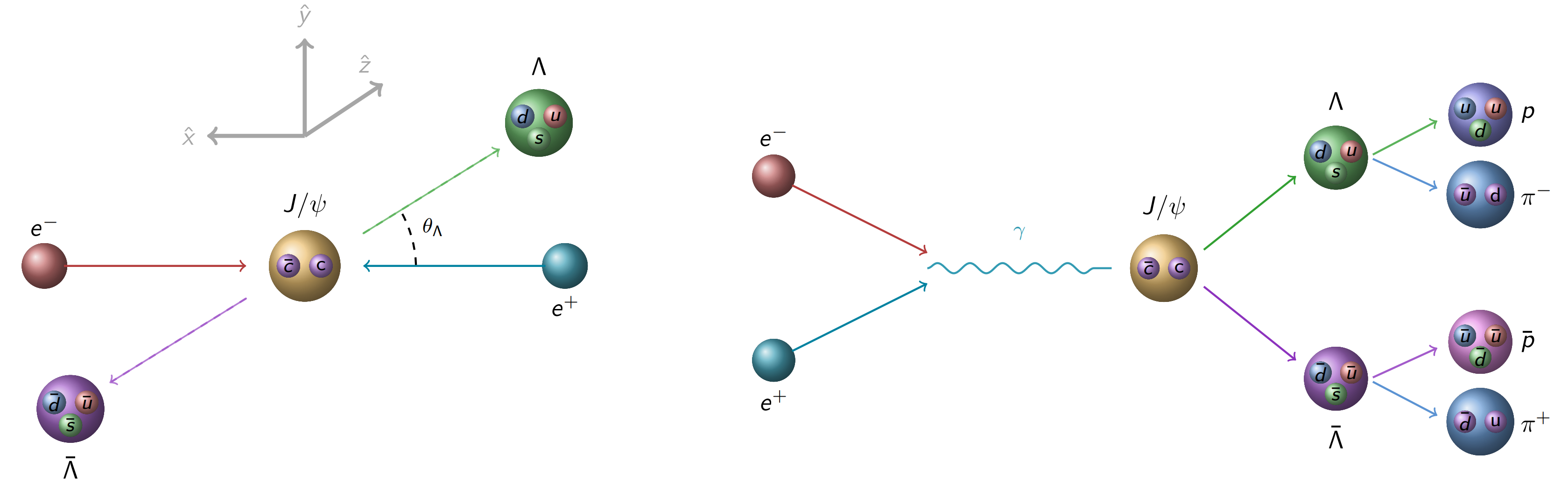}
\end{center}
\caption{The processes of
$e^++e^-\to J/\psi\to \Lambda(\to p+\pi^-)+\bar{\Lambda}(\to \bar{p}+\pi^+)$. 
}\label{decay}
\end{figure}

Fig.~\ref{decay} shows the $\Lambda \bar{\Lambda}$ production through $J/\psi$ decay in electron-positron collisions, followed by the $\Lambda$ and $\bar{\Lambda}$ decay processes given in Eq.~(\ref{llde}):  
\begin{align}  
e^++e^-\to J/\psi\to \Lambda(\to p+\pi^-)+\bar{\Lambda}(\to \bar{p}+\pi^+)~.  \label{caa}
\end{align}  
The $J/\psi-\Lambda\bar{\Lambda}$ interaction is conventionally parameterized as \cite{Dalkarov:2009yf,Czyz:2007wi,Faldt:2013gka}:  
\begin{align}  
   & \psi_\mu \bar{\Lambda} \left(G_M\gamma^\mu+\frac{2m_\Lambda}{m_\psi^2-4m_\Lambda^2}\left(G_M-G_E\right)\left(p_\Lambda^\mu-p_{\bar{\Lambda}}^\mu\right)\right)\Lambda~,\\  
  & G_E/G_M=R e^{i\Delta\phi}~.  
\end{align}  
In the $J/\psi$ rest frame, the $\Lambda$ polar angle distribution follows:  
\begin{align}  
   & \frac{1}{\Gamma_{\psi \to \Lambda+\bar{\Lambda}}}\frac{d\Gamma_{\psi \to \Lambda+\bar{\Lambda}}}{d\cos\theta_\Lambda}=\frac{1}{2+\frac{2}{3}\alpha_\psi}\left(1+\alpha_\psi \cos^2\theta_\Lambda\right)~,  
\end{align}  
with asymmetry parameter:  
\begin{align}  
   & \alpha_\psi=\left(1-4\frac{m_\Lambda^2}{m_\psi^2}R^2\right)/\left(1+4\frac{m_\Lambda^2}{m_\psi^2}R^2\right)~.
\end{align}  
Using almost equal decay widths for \(\Lambda\) and \(\bar{\Lambda}\) \cite{ParticleDataGroup:2024cfk}, and applying Eq.~(\ref{spcl}) to the \( J/\psi\to \Lambda+\bar{\Lambda} \) process, we derive:  
\begin{align}  
    \mathcal{P}(\Delta s^2<0)=\sqrt{1-4m_\Lambda^2/m_\psi^2}~.  
\end{align}  
{So, for a fermion-antifermion decay system, the probability that the decay of the fermion and that of the antifermion are space-like separated is equal to their speed in the center-of-mass frame (in units where the speed of light $c$ is set to unity).
This result implies that \(69.3\%\) of the decay events involving \(\Lambda\to p+\pi^-\) and \(\bar{\Lambda}\to \bar{p}+\pi^+\) are spacelike-separated.
It is instructive to contrast this with the \(t\bar t\) system: while entanglement in \(t\bar t\) production is most pronounced near the threshold region \(\sqrt{s}\sim 2m_t\) \cite{ATLAS2024,CMS:2024pts}, the corresponding boost factor there is small, which significantly reduces the fraction of spacelike-separated decays in that same kinematic regime.
}

We define
\begin{align}
    \Delta_{\Lambda\bar{\Lambda}}=\alpha_{\frac{1}{2},\frac{1}{2}}\alpha_{-\frac{1}{2},-\frac{1}{2}}-\alpha_{\frac{1}{2},-\frac{1}{2}}\alpha_{-\frac{1}{2},\frac{1}{2}}~.
\end{align}
The necessary and sufficient condition for quantum entanglement in the $\Lambda\bar{\Lambda}$ system in a pure state is given by \cite{Pei:2025non}  
\begin{align}  
    \Delta_{\Lambda\bar{\Lambda}}\ne0~.  
\end{align}  
For $\Lambda\bar{\Lambda}$ pairs produced through $e^+ +e^-\to J/\psi\to \Lambda+\bar{\Lambda}$, theoretical calculations under decoherence-free conditions yield:  
\begin{align}  
    \Delta_{\Lambda\bar{\Lambda}}=\frac{1-\alpha_\psi}{4}\frac{1-\cos^2\theta_\Lambda}{1+\alpha_\psi\cos^2\theta_\Lambda}\left(\frac{1+\alpha_\psi}{1-\alpha_\psi}-e^{2i\Delta\phi}\right)~.  
\end{align}    
This implies that under decoherence-free evolution, quantum entanglement persists for all $\Lambda\bar{\Lambda}$ systems with $-1<\cos\theta_\Lambda<1$, unless both $\alpha_\psi$ and $\Delta\phi$ vanish simultaneously.
However, the quantity $\Delta_{\Lambda\bar{\Lambda}}$ does not directly correspond to any single physical observable. As established in Ref.~\cite{Pei:2025non}, the magnitude $\left|\Delta_{\Lambda\bar{\Lambda}}\right|$ can be derived through complex algebraic operations involving statistical averages of multiple physical observables. Consequently, experimental determination of $\left|\Delta_{\Lambda\bar{\Lambda}}\right|$ is not straightforward.

The complete angular distribution for the cascade decay in Eq.~(\ref{caa}) process satisfies \cite{Faldt:2017kgy}
\begin{align}  
    \mathcal{W}=&\mathcal{T}_0+\alpha_\psi\mathcal{T}_5\nonumber\\  
    &+\alpha_\Lambda \alpha_{\bar{\Lambda}}\left(\mathcal{T}_1+\sqrt{1-\alpha_\psi^2}\cos(\Delta\phi)\mathcal{T}_2+\alpha_\psi\mathcal{T}_6\right)\nonumber\\  
    &+\sqrt{1-\alpha_\psi^2}\sin(\Delta\phi)\left(\alpha_\Lambda \mathcal{T}_3+\alpha_{\bar{\Lambda}} \mathcal{T}_4\right)~,  
\end{align}  
where angular variables $\theta_1$, $\theta_2$, $\phi_1$, $\phi_2$ follow definitions in Eqs.~(\ref{jiao1}) and (\ref{jiao2}), with  
\begin{align}  
    &\mathcal{T}_0=1~,\\  
    &\mathcal{T}_1=\sin\theta_\Lambda^2\sin\theta_1 \sin\theta_2\cos\phi_1 \cos\phi_2+\cos^2\theta_\Lambda\cos\theta_1\cos\theta_2~,\\  
    &\mathcal{T}_2= \sin\theta_\Lambda \cos\theta_\Lambda\left(\sin\theta_1 \cos\theta_2\cos\phi_1+\cos\theta_1\sin\theta_2\cos\phi_2\right)~,\\  
  &\mathcal{T}_3= \sin\theta_\Lambda \cos\theta_\Lambda\sin\theta_1 \sin\phi_1~,\\  
&\mathcal{T}_4= \sin\theta_\Lambda \cos\theta_\Lambda\sin\theta_2 \sin\phi_2~,\\  
&\mathcal{T}_5=\cos^2\theta_\Lambda~,\\  
&\mathcal{T}_6=\cos\theta_1\cos\theta_2-\sin^2\theta_\Lambda \sin\theta_1\sin\theta_2\sin\phi_1\sin\phi_2~.  
\end{align}  
Direct integration yields
\begin{align}
    \langle\mathcal{O}(\theta_1,\theta_2,\phi_1,\phi_2)\rangle=\frac{1}{N}\int \mathcal{O}(\theta_1,\theta_2,\phi_1,\phi_2)~\mathcal{W}~d\Omega_p ~d\Omega_{\bar{p}}
\end{align}
with
\begin{align}
   & N=\int\mathcal{W}~d\Omega_p ~d\Omega_{\bar{p}}~,\quad d\Omega_{p/\bar{p}}=d\cos\theta_{1/2} d\phi_{1/2}~.
\end{align}
This leads to the normalized observables:  
\begin{align}  
    \mathcal{O}_0&= -1~,\\
    \mathcal{O}_1&=- \frac{1}{2} (1+\alpha_\psi)\frac{1-\cos^2\theta_\Lambda}{1+\alpha_\psi\cos^2\theta_\Lambda}~,\\  
    \mathcal{O}_2&=- \frac{1}{2}(1-\alpha_\psi)\frac{1-\cos^2\theta_\Lambda}{1+\alpha_\psi\cos^2\theta_\Lambda}~,\\
    \mathcal{O}_3&=0~,\\  
    \mathcal{O}_4&=0~.
\end{align}  
So, $\mathcal{O}_0$, $\mathcal{O}_3$, and $\mathcal{O}_4$ all do not violate their respective product state boundaries, thus can not demonstrate quantum entanglement in $\Lambda\bar{\Lambda}$.

At $\cos\theta_\Lambda=0$, the observables attain their minimal values:  
\begin{align}  
    &\mathcal{O}_{1\text{min}}=-\frac{1+\alpha_\psi}{2}~,\\  
    &\mathcal{O}_{2\text{min}}=-\frac{1-\alpha_\psi}{2}~,
\end{align}    
and when $0<\alpha_\psi<1$,  the correlation parameter $\mathcal{O}_1$ satisfies:  
\begin{align}  
   \mathcal{O}_{1}<-\frac{1}{2}~,\quad \text{where}~ \left|\cos\theta_\Lambda\right|<1/\sqrt{2+1/\alpha_\psi}~.  
\end{align}  

\begin{figure}[tbp]
\begin{center}
\includegraphics[width=0.9\textwidth]{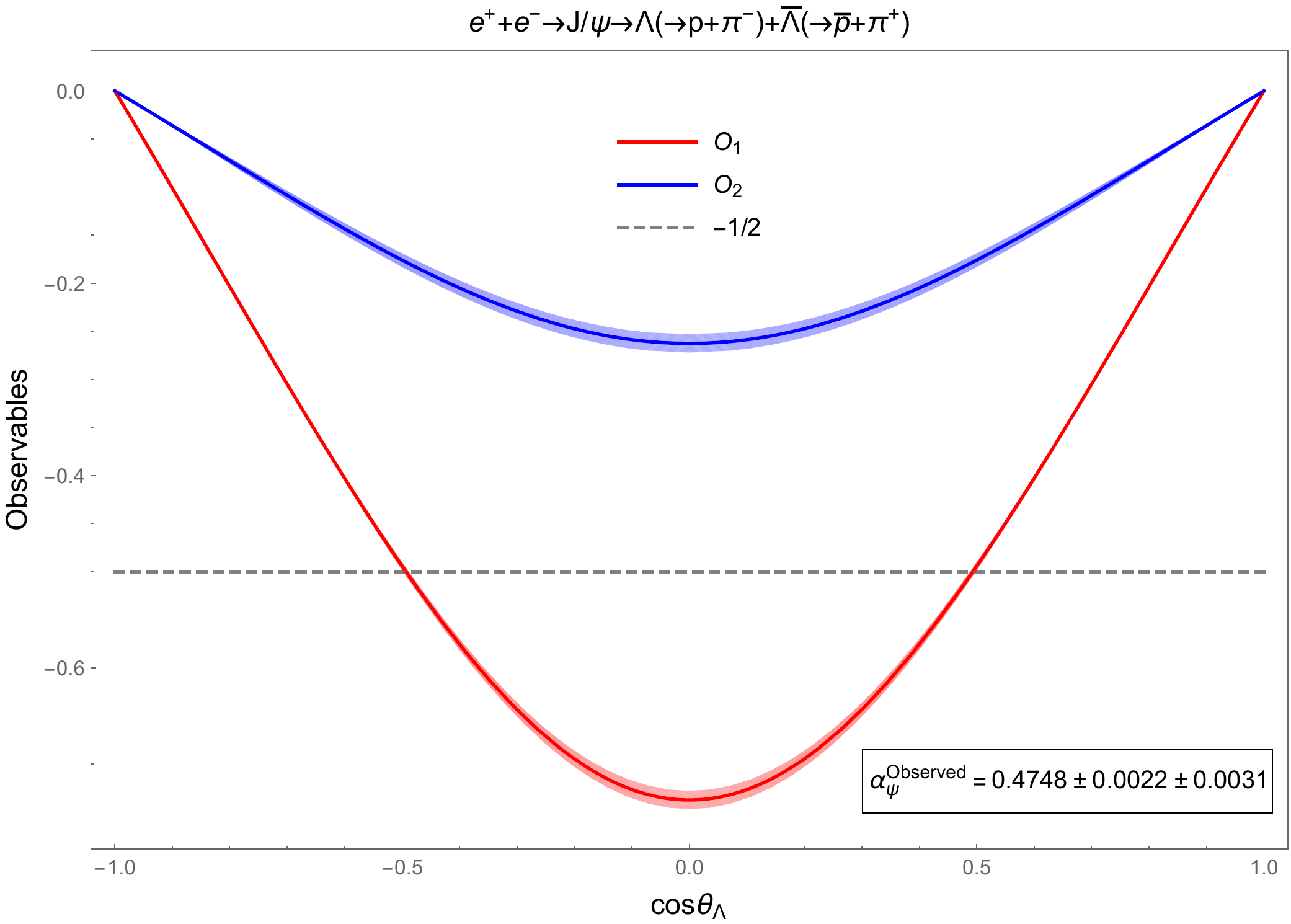}
\end{center}
 \vspace*{-0.1in}
\caption{The $\mathcal{O}_1=\langle \cos(\phi_1+\phi_2)\rangle/\left({-\frac{\pi^2}{32}\alpha_\Lambda \alpha_{\bar{\Lambda}}}\right)$ and $\mathcal{O}_2={\langle \cos(\phi_1-\phi_2)\rangle}/\left({-\frac{\pi^2}{32}\alpha_\Lambda \alpha_{\bar{\Lambda}}}\right)$ for \( \Lambda\bar{\Lambda} \) pairs produced at \( e^+ e^- \) collider by using $ \alpha_\psi^{\text{Observed}}=0.4748\pm 0.0022\pm 0.0031$ \cite{BESIII:2022qax}. The solid lines are given by the central value of $\alpha_\psi^{\text{Observed}}$, while the shaded bands correspond to the 5$\sigma$ confidence region of $\alpha_\psi^{\text{Observed}}$.
}\label{dp}
\end{figure}

The measured value of $\alpha_\psi$ is determined as \cite{BESIII:2022qax}:  
\begin{align}  
    \alpha_\psi^{\text{Observed}}=0.4748\pm 0.0022\pm 0.0031~. 
\end{align}  
Fig.~\ref{dp} displays the normalized correlation observables $\mathcal{O}_1$ and $\mathcal{O}_2$ for $ \Lambda\bar{\Lambda} $ pairs produced at $ e^+ e^- $ colliders by using  $\alpha_\psi^{\text{Observed}}$.
The solid lines are  $\mathcal{O}_i~(i=1,2)$ given by the central value of $\alpha_\psi^{\text{Observed}}$, while the shaded bands correspond to the 5$\sigma$ confidence region of $\alpha_\psi^{\text{Observed}}$.
It is shown that the observed minimum value of $\mathcal{O}_1$ is
\begin{align}  
    \mathcal{O}_{1\text{min}}^{\text{Observed}}=-0.7374\pm 0.0011\pm 0.0016~, 
\end{align}    
{which significantly exceeding the classical limit of \(-\frac{1}{2}\). {This measurement thus experimentally validates the polar–azimuthal selection rule: the violation of the product-state bound is observed precisely in the azimuthal coherence channel.} Furthermore, for $\left|\cos\theta_\Lambda\right|<0.4883$, the observed $\mathcal{O}_{1}^{\text{Observed}}$ consistently exhibits $\mathcal{O}_{1}^{\text{Observed}} < -\frac{1}{2}$ with a statistical significance of at least 5$\sigma$.
To assess the robustness of the present analysis, we follow the uncertainty treatment adopted in Ref.~\cite{BESIII:2022qax}. In that work, the systematic uncertainties were evaluated from several independent sources, namely the background description, tracking efficiency, the \(\Lambda/\bar{\Lambda}\) vertex fit, the kinematic fit, and the fit method itself. Their absolute systematic uncertainties, summarized in Table II of Ref. \cite{BESIII:2022qax}, are all at the \(10^{-3}\) level: for example, the background uncertainty is \(0.0\text{--}0.4\times 10^{-3}\), the tracking uncertainty is \(0.2\text{--}2.1\times 10^{-3}\), the \(\Lambda/\bar{\Lambda}\) vertex-fit uncertainty is \(0.0\text{--}0.3\times 10^{-3}\), the kinematic-fit uncertainty is \(0.3\text{--}3.0\times 10^{-3}\), and the fit-method uncertainty is \(1.0\text{--}5.8\times 10^{-3}\), depending on the fitted parameter. The total systematic uncertainty is obtained by adding these contributions in quadrature, yielding values of only a few parts in \(10^{-3}\) for the measured observables. In particular, even for the parameter with the largest quoted systematic effect, the total uncertainty remains at the \(10^{-3}\) level. Since our analysis is based directly on the published measurements and fit results of Ref. \cite{BESIII:2022qax}, the relevant experimental uncertainties are already encoded in the input data. Propagating these uncertainties through our observables induces only a tiny variation, far smaller than the central effect we discuss. So, the experimentally determined spin-correlation observable $\mathcal{O}_1$ for the $\Lambda\bar{\Lambda}$ system lie well within the entangled region, providing a high-precision characterization of entanglement in this channel.
}

{Before closing this section, we would like to highlight several additional conceptual advantages of the present measurement. First, the \(\Lambda(\to p\pi^-)\bar{\Lambda}(\to \bar{p}\pi^+)\) system provides a more generic setting for entanglement studies of fermion-antifermion pairs. In existing studies of entanglement in \(t(\to e^+ \nu_e b)\bar{t}(\to e^- \bar{\nu}_e \bar{b})\) \cite{ATLAS2024,CMS:2024pts}, the corresponding spin-analyzing powers take the extreme values \(\alpha_t=1\) and \(\alpha_{\bar t}=-1\). Since spin-analyzing powers are generally bounded within the interval \([-1,1]\), these endpoint values are highly special. Consequently, one may be led to the misleading impression that the entanglement properties of this system are insensitive to the specific values of the spin-analyzing powers. By contrast, in the \(\Lambda(\to p\pi^-)\bar{\Lambda}(\to \bar{p}\pi^+)\) channel, the spin-analyzing powers take more generic, non-extremal values. In this sense, the \(\Lambda\bar{\Lambda}\) system provides a more representative and conceptually more general framework for investigating how entanglement manifests itself in realistic fermion-antifermion systems.

Second, in \(e^+e^- \to J/\psi \to \Lambda\bar\Lambda\), clear quantum entanglement is observed in the region \(\cos\theta_\Lambda \simeq 0\), while a relatively high fraction of spacelike-separated events is maintained, in contrast to the \(t\bar t\) system produced at the LHC \cite{ATLAS2024,CMS:2024pts}. Thus, the advantage of the \(\Lambda\bar\Lambda\) channel lies not only in its pronounced sensitivity to entanglement itself, but also in its ability to combine strong entanglement signals with a high efficiency of spacelike-separated configurations. Although the present work does not address Bell nonlocality, this distinctive combination makes the system particularly promising for future studies aimed at probing nonlocal features in high-energy collisions.

Third, because the full kinematics of the process can be reconstructed on an event-by-event basis, the \(\Lambda\bar\Lambda\) system allows us to construct a broad class of experimentally accessible observables and, as shown in our work, to derive the corresponding entanglement criteria for each of them. Therefore, this channel is not limited to a single entanglement test quantity; rather, it provides a diverse set of measurable observables equipped with explicit separability bounds. This richness is particularly important because, for different \(\Lambda\bar\Lambda\) production mechanisms, the most sensitive observable and the most effective entanglement criterion need not be the same. Having a sufficiently large set of pre-constructed observables and criteria greatly enhances both the efficiency and the adaptability of the entanglement search, allowing one to choose the most suitable probe for a given production process or experimental condition. We believe that this flexibility and diversity constitute a distinctive advantage of the \(\Lambda\bar\Lambda\) channel and help clarify its added value relative to other systems.
}

\section{Conclusions}\label{sec:con}

{This work provides strong experimental evidences for QE in the \(\Lambda\bar{\Lambda}\) system produced through \(e^+e^-\) annihilation at the \(J/\psi\) resonance.} By constructing a complete theoretical framework for spin correlations, we identified several key observables that distinguish entangled states from product ones:  \(\mathcal{O}_0 = \langle \hat{e}_p\cdot \hat{e}_{\bar{p}}\rangle/C_0\), \(\mathcal{O}_1 = \langle \cos(\phi_1+\phi_2) \rangle / C_1\), \(\mathcal{O}_2 = \langle \cos(\phi_1-\phi_2) \rangle / C_1\), \(\mathcal{O}_3 = \langle \sin(\phi_1+\phi_2) \rangle / C_1\), and \(\mathcal{O}_4 = \langle \sin(\phi_1-\phi_2) \rangle / C_1\), where $C_0=-\frac{1}{9}\alpha_\Lambda \alpha_{\bar{\Lambda}}$ and $C_1=-\frac{\pi^2}{32}\alpha_\Lambda \alpha_{\bar{\Lambda}}$. 

{The measurement of \(\mathcal{O}_{1\text{min}}^\text{Observed} = -0.7374\pm 0.0011\pm 0.0016\) at \(\cos\theta_\Lambda = 0\) clearly violates the product state boundary of \([-0.5,0.5]\). For $\left|\cos\theta_\Lambda\right|<0.4883$, the observed $\mathcal{O}_{1}^{\text{Observed}}$ consistently exhibits $\mathcal{O}_{1}^{\text{Observed}} < -\frac{1}{2}$ with a statistical significance of at least 5$\sigma$.} These constitute the first observation of QE in a hyperon-antihyperon system. This result is particularly significant because it demonstrates QE persists through both the strong interaction production process (\(J/\psi \to \Lambda\bar{\Lambda}\)) and weak decay processes (\(\Lambda \to p+\pi^-\) and \(\bar{\Lambda} \to \bar{p}+\pi^+\)).

{The novelty of this work is twofold. First, it provides a fully reconstructible, clean, and generic-parameter benchmark for entanglement in a baryon–antibaryon system, complementing top-quark and tau-lepton studies. Second, it establishes a general polar–azimuthal selection rule: polar-angle-only observables cannot certify entanglement, whereas azimuthal correlations are necessary.}
Our analysis method, based on the complete angular distribution, provides a template for future QE studies in other baryon-antibaryon systems. The success of this approach suggests several promising directions:  
\begin{itemize}
    \item Extension to other production channels like \(\psi(2S) \to \Lambda\bar{\Lambda}\)
    \item Application to spin-3/2 hyperons like \(\Omega^-\bar{\Omega}^+\)
    \item Investigation of CP violation through QE observables
    \item Tests of decoherence effects in high-energy collisions 
\end{itemize}
These results establish hyperon pairs as a new laboratory for {measuring quantum properties} in relativistic systems, complementing traditional tests with photons or atoms. The demonstrated persistence of QE through strong and weak interactions opens new possibilities for fundamental explorations at particle colliders.  

\begin{acknowledgments}

J. Pei is supported by the National Natural Science Foundation of China (Project No. 12505121), by the Joint Fund of Henan Province Science and Technology R$\&$D Program (Project No. 245200810077), by the Startup Research Fund of Henan Academy of Sciences (Project No. 20251820001),
and by the Scientific and Technological Research Project of Henan Academy of Sciences (Project No. 20262320001).
TL is supported in part by the National Key Research and Development Program of China Grant No. 2020YFC2201504, by the
Projects No. 11875062, No. 11947302, No. 12047503,
and No. 12275333 supported by the National Natural Science Foundation of China, by the Key Research
Program of the Chinese Academy of Sciences, Grant
No. XDPB15, by the Scientific Instrument Developing Project of the Chinese Academy of Sciences, Grant
No. YJKYYQ20190049, by the International Partner
ship Program of Chinese Academy of Sciences for Grand
Challenges, Grant No. 112311KYSB20210012, and by
the Henan Province Outstanding Foreign Scientist Studio Project, No.GZS2025008.

\end{acknowledgments}

\bibliographystyle{jhep}
\bibliography{jhep}

\providecommand{\href}[2]{#2}\begingroup\raggedright\begin{thebibliography}{10}

\bibitem{PhysicsPhysiqueFizika.1.195}
J.S.~Bell, \emph{On the einstein podolsky rosen paradox}, \href{https://doi.org/10.1103/PhysicsPhysiqueFizika.1.195}{\emph{Physics Physique Fizika} {\bfseries 1} (1964) 195}.

\bibitem{PhysRevLett.28.938}
S.J.~Freedman and J.F.~Clauser, \emph{Experimental test of local hidden-variable theories}, \href{https://doi.org/10.1103/PhysRevLett.28.938}{\emph{Phys. Rev. Lett.} {\bfseries 28} (1972) 938}.

\bibitem{PhysRevLett.47.460}
A.~Aspect, P.~Grangier and G.~Roger, \emph{Experimental tests of realistic local theories via bell's theorem}, \href{https://doi.org/10.1103/PhysRevLett.47.460}{\emph{Phys. Rev. Lett.} {\bfseries 47} (1981) 460}.

\bibitem{PhysRevLett.49.1804}
A.~Aspect, J.~Dalibard and G.~Roger, \emph{Experimental test of bell's inequalities using time-varying analyzers}, \href{https://doi.org/10.1103/PhysRevLett.49.1804}{\emph{Phys. Rev. Lett.} {\bfseries 49} (1982) 1804}.

\bibitem{PhysRevLett.81.3563}
W.~Tittel, J.~Brendel, H.~Zbinden and N.~Gisin, \emph{Violation of bell inequalities by photons more than 10 km apart}, \href{https://doi.org/10.1103/PhysRevLett.81.3563}{\emph{Phys. Rev. Lett.} {\bfseries 81} (1998) 3563}.

\bibitem{Hensen:2015ccp}
B.~Hensen et~al., \emph{{Loophole-free Bell inequality violation using electron spins separated by 1.3 kilometres}}, \href{https://doi.org/10.1038/nature15759}{\emph{Nature} {\bfseries 526} (2015) 682} [\href{https://arxiv.org/abs/1508.05949}{{\ttfamily 1508.05949}}].

\bibitem{PhysRevLett.70.1895}
C.H.~Bennett, G.~Brassard, C.~Cr\'epeau, R.~Jozsa, A.~Peres and W.K.~Wootters, \emph{Teleporting an unknown quantum state via dual classical and einstein-podolsky-rosen channels}, \href{https://doi.org/10.1103/PhysRevLett.70.1895}{\emph{Phys. Rev. Lett.} {\bfseries 70} (1993) 1895}.

\bibitem{Bouwmeester_1997}
D.~Bouwmeester, J.-W.~Pan, K.~Mattle, M.~Eibl, H.~Weinfurter and A.~Zeilinger, \emph{Experimental quantum teleportation}, \href{https://doi.org/10.1038/37539}{\emph{Nature} {\bfseries 390} (1997) 575–579}.

\bibitem{PhysRevLett.80.1121}
D.~Boschi, S.~Branca, F.~De~Martini, L.~Hardy and S.~Popescu, \emph{Experimental realization of teleporting an unknown pure quantum state via dual classical and einstein-podolsky-rosen channels}, \href{https://doi.org/10.1103/PhysRevLett.80.1121}{\emph{Phys. Rev. Lett.} {\bfseries 80} (1998) 1121}.

\bibitem{Riebe:2004jpa}
M.~Riebe et~al., \emph{{Deterministic quantum teleportation with atoms}}, \href{https://doi.org/10.1038/nature02570}{\emph{Nature} {\bfseries 429} (2004) 734}.

\bibitem{Barrett:2004bxh}
M.D.~Barrett et~al., \emph{{Deterministic quantum teleportation of atomic qubits}}, \href{https://doi.org/10.1038/nature02608}{\emph{Nature} {\bfseries 429} (2004) 737}.

\bibitem{vonKuk:2025kbv}
R.~von Kuk, K.~Lee, J.K.L.~Michel and Z.~Sun, \emph{{Towards a Quantum Information Theory of Hadronization: Dihadron Fragmentation and Neutral Polarization in Heavy Baryons}},  \href{https://arxiv.org/abs/2503.22607}{{\ttfamily 2503.22607}}.

\bibitem{Subba:2024mnl}
A.~Subba and R.~Rahaman, \emph{{On bipartite and tripartite entanglement at present and future particle colliders}},  \href{https://arxiv.org/abs/2404.03292}{{\ttfamily 2404.03292}}.

\bibitem{Han:2025ewp}
T.~Han, M.~Low and Y.~Su, \emph{{Entanglement and Bell Nonlocality in $\tau^+ \tau^-$ at the BEPC}},  \href{https://arxiv.org/abs/2501.04801}{{\ttfamily 2501.04801}}.

\bibitem{Afik:2020onf}
Y.~Afik and J.R.M.n.~de~Nova, \emph{{Entanglement and quantum tomography with top quarks at the LHC}}, \href{https://doi.org/10.1140/epjp/s13360-021-01902-1}{\emph{Eur. Phys. J. Plus} {\bfseries 136} (2021) 907} [\href{https://arxiv.org/abs/2003.02280}{{\ttfamily 2003.02280}}].

\bibitem{Fabbrichesi:2021npl}
M.~Fabbrichesi, R.~Floreanini and G.~Panizzo, \emph{{Testing Bell Inequalities at the LHC with Top-Quark Pairs}}, \href{https://doi.org/10.1103/PhysRevLett.127.161801}{\emph{Phys. Rev. Lett.} {\bfseries 127} (2021) 161801} [\href{https://arxiv.org/abs/2102.11883}{{\ttfamily 2102.11883}}].

\bibitem{Han:2023fci}
T.~Han, M.~Low and T.A.~Wu, \emph{{Quantum entanglement and Bell inequality violation in semi-leptonic top decays}}, \href{https://doi.org/10.1007/JHEP07(2024)192}{\emph{JHEP} {\bfseries 07} (2024) 192} [\href{https://arxiv.org/abs/2310.17696}{{\ttfamily 2310.17696}}].

\bibitem{ATLAS2024}
{ATLAS Collaboration}, \emph{Observation of quantum entanglement with top quarks at the {ATLAS} detector}, {\emph{Nature} {\bfseries 633} (2024) 542} [\href{https://arxiv.org/abs/2311.07288}{{\ttfamily 2311.07288}}].

\bibitem{Cheng:2023qmz}
K.~Cheng, T.~Han and M.~Low, \emph{{Optimizing fictitious states for Bell inequality violation in bipartite qubit systems with applications to the tt{\textasciimacron} system}}, \href{https://doi.org/10.1103/PhysRevD.109.116005}{\emph{Phys. Rev. D} {\bfseries 109} (2024) 116005} [\href{https://arxiv.org/abs/2311.09166}{{\ttfamily 2311.09166}}].

\bibitem{CMS:2024pts}
{\scshape CMS} Collaboration, \emph{{Observation of quantum entanglement in top quark pair production in proton\textendash{}proton collisions at $\sqrt{s} = 13$ TeV}}, \href{https://doi.org/10.1088/1361-6633/ad7e4d}{\emph{Rept. Prog. Phys.} {\bfseries 87} (2024) 117801} [\href{https://arxiv.org/abs/2406.03976}{{\ttfamily 2406.03976}}].

\bibitem{Cheng:2025cuv}
K.~Cheng and B.~Yan, \emph{{Bell Inequality Violation of Light Quarks in Dihadron Pair Production at Lepton Colliders}}, \href{https://doi.org/10.1103/gmqz-v4cl}{\emph{Phys. Rev. Lett.} {\bfseries 135} (2025) 011902} [\href{https://arxiv.org/abs/2501.03321}{{\ttfamily 2501.03321}}].

\bibitem{Subba:2024aut}
A.~Subba, R.K.~Singh and R.M.~Godbole, \emph{{Looking into the quantum entanglement in $H\to ZZ^\star$ at LHC within SMEFT framework}},  \href{https://arxiv.org/abs/2411.19171}{{\ttfamily 2411.19171}}.

\bibitem{BESIII:2018cnd}
{\scshape BESIII} Collaboration, \emph{{Polarization and Entanglement in Baryon-Antibaryon Pair Production in Electron-Positron Annihilation}}, \href{https://doi.org/10.1038/s41567-019-0494-8}{\emph{Nature Phys.} {\bfseries 15} (2019) 631} [\href{https://arxiv.org/abs/1808.08917}{{\ttfamily 1808.08917}}].

\bibitem{BESIII:2025vsr}
{\scshape BESIII} Collaboration, \emph{{Test of local realism via entangled $\Lambda\bar\Lambda$ system}},  \href{https://arxiv.org/abs/2505.14988}{{\ttfamily 2505.14988}}.

\bibitem{Lin:2025eci}
S.-J.~Lin, M.-J.~Liu, D.Y.~Shao and S.-Y.~Wei, \emph{{Spin correlations and Bell nonlocality in $\Lambda\bar{\Lambda}$ pair production from $e^+e^-$ collisions with a thrust cut}},  \href{https://arxiv.org/abs/2507.15387}{{\ttfamily 2507.15387}}.

\bibitem{Wu:2025dds}
S.~Wu, C.~Qian, Q.~Wang and Y.-G.~Yang, \emph{{Quantum steering and discord in hyperon-antihyperon system in electron-positron annihilation}},  \href{https://arxiv.org/abs/2509.14990}{{\ttfamily 2509.14990}}.

\bibitem{ABEL1992304}
S.~Abel, M.~Dittmar and H.~Dreiner, \emph{Testing locality at colliders via bell's inequality?}, \href{https://doi.org/https://doi.org/10.1016/0370-2693(92)90071-B}{\emph{Physics Letters B} {\bfseries 280} (1992) 304}.

\bibitem{Li:2024luk}
S.~Li, W.~Shen and J.M.~Yang, \emph{{Can Bell inequalities be tested via scattering cross-section at colliders ?}}, \href{https://doi.org/10.1140/epjc/s10052-024-13584-x}{\emph{Eur. Phys. J. C} {\bfseries 84} (2024) 1195} [\href{https://arxiv.org/abs/2401.01162}{{\ttfamily 2401.01162}}].

\bibitem{Bechtle:2025ugc}
P.~Bechtle, C.~Breuning, H.K.~Dreiner and C.~Duhr, \emph{{A critical appraisal of tests of locality and of entanglement versus non-entanglement at colliders}},  \href{https://arxiv.org/abs/2507.15947}{{\ttfamily 2507.15947}}.

\bibitem{Abel:2025skj}
S.A.~Abel, H.K.~Dreiner, R.~Sengupta and L.~Ubaldi, \emph{{Colliders are Testing neither Locality via Bell's Inequality nor Entanglement versus Non-Entanglement}},  \href{https://arxiv.org/abs/2507.15949}{{\ttfamily 2507.15949}}.

\bibitem{Pei:2026wfu}
J.~Pei, L.~Wu, T.~Li and X.~Hao, \emph{{Excluding Local Hidden Variables in $\Lambda\bar{\Lambda}$ Production: The Incompatibility with Angular-Momentum Conservation and CPT Invariance}},  \href{https://arxiv.org/abs/2601.15747}{{\ttfamily 2601.15747}}.

\bibitem{Pei:2026rlh}
J.~Pei, L.~Wu, D.~Wang, X.~Hao and T.~Li, \emph{{Unveiling a Universal Formalism for Quantum Entanglement in Arbitrary Spin Decays}},  \href{https://arxiv.org/abs/2601.15748}{{\ttfamily 2601.15748}}.

\bibitem{Pei:2025non}
J.~Pei, Y.~Fang, L.~Wu, D.~Xu, M.~Biyabi and T.~Li, \emph{{Quantum Entanglement Theory and Its Generic Searches in High Energy Physics}},  \href{https://arxiv.org/abs/2505.09280}{{\ttfamily 2505.09280}}.

\bibitem{Faldt:2013gka}
G.~F\"aldt, \emph{{Entanglement in joint $\Lambda \bar{\Lambda}$ decay}}, \href{https://doi.org/10.1140/epja/i2015-15074-3}{\emph{Eur. Phys. J. A} {\bfseries 51} (2015) 74} [\href{https://arxiv.org/abs/1306.0525}{{\ttfamily 1306.0525}}].

\bibitem{BESIII:2021ypr}
{\scshape BESIII} Collaboration, \emph{{Probing CP symmetry and weak phases with entangled double-strange baryons}}, \href{https://doi.org/10.1038/s41586-022-04624-1}{\emph{Nature} {\bfseries 606} (2022) 64} [\href{https://arxiv.org/abs/2105.11155}{{\ttfamily 2105.11155}}].

\bibitem{Scherer:2002tk}
S.~Scherer, \emph{{Introduction to chiral perturbation theory}}, {\emph{Adv. Nucl. Phys.} {\bfseries 27} (2003) 277} [\href{https://arxiv.org/abs/hep-ph/0210398}{{\ttfamily hep-ph/0210398}}].

\bibitem{Leader:2001nas}
E.~Leader, \emph{{Spin in Particle Physics}}, vol.~15, Cambridge University Press (2001), \href{https://doi.org/10.1017/9781009402040}{10.1017/9781009402040}.

\bibitem{Pei:2025ito}
J.~Pei, T.~Li, L.~Wu, X.~Hao and X.~Wang, \emph{{Bypassing spin-analyzing power dependence for quantum entanglement at colliders: a case study of $\Lambda \overline{\Lambda}$}}, \href{https://doi.org/10.1007/JHEP07(2026)033}{\emph{JHEP} {\bfseries 07} (2026) 33} [\href{https://arxiv.org/abs/2510.08031}{{\ttfamily 2510.08031}}].

\bibitem{Dalkarov:2009yf}
O.D.~Dalkarov, P.A.~Khakhulin and A.Y.~Voronin, \emph{{On the electromagnetic form factors of hadrons in the time-like region near threshold}}, \href{https://doi.org/10.1016/j.nuclphysa.2009.11.015}{\emph{Nucl. Phys. A} {\bfseries 833} (2010) 104} [\href{https://arxiv.org/abs/0906.0266}{{\ttfamily 0906.0266}}].

\bibitem{Czyz:2007wi}
H.~Czyz, A.~Grzelinska and J.H.~Kuhn, \emph{{Spin asymmetries and correlations in lambda-pair production through the radiative return method}}, \href{https://doi.org/10.1103/PhysRevD.75.074026}{\emph{Phys. Rev. D} {\bfseries 75} (2007) 074026} [\href{https://arxiv.org/abs/hep-ph/0702122}{{\ttfamily hep-ph/0702122}}].

\bibitem{ParticleDataGroup:2024cfk}
{\scshape Particle Data Group} Collaboration, \emph{{Review of particle physics}}, \href{https://doi.org/10.1103/PhysRevD.110.030001}{\emph{Phys. Rev. D} {\bfseries 110} (2024) 030001}.

\bibitem{Faldt:2017kgy}
G.~F\"aldt and A.~Kupsc, \emph{{Hadronic structure functions in the $e^+ e^- \rightarrow \bar{\Lambda} \Lambda$ reaction}}, \href{https://doi.org/10.1016/j.physletb.2017.06.011}{\emph{Phys. Lett. B} {\bfseries 772} (2017) 16} [\href{https://arxiv.org/abs/1702.07288}{{\ttfamily 1702.07288}}].

\bibitem{BESIII:2022qax}
{\scshape BESIII} Collaboration, \emph{{Precise Measurements of Decay Parameters and $CP$ Asymmetry with Entangled $\Lambda-\bar{\Lambda}$ Pairs}}, \href{https://doi.org/10.1103/PhysRevLett.129.131801}{\emph{Phys. Rev. Lett.} {\bfseries 129} (2022) 131801} [\href{https://arxiv.org/abs/2204.11058}{{\ttfamily 2204.11058}}].

\end{thebibliography}\endgroup

\end{document}